\documentclass[review
]{elsarticle}
\usepackage[cp1251]{inputenc}
\usepackage[english
]{babel}%

\usepackage{amsmath}
\usepackage{latexsym}
\usepackage{amsbsy}
\usepackage{amsfonts}
\usepackage{amssymb}
\usepackage{euscript}
\usepackage{xcolor}
\usepackage{graphicx}
\usepackage{lineno,hyperref}
\usepackage{txfonts}

\def \R {{\bf R}}
\def \r {{\bf r}}
\def \u {{\bf u}}

\begin{document}

\begin{frontmatter}

\title{Modeling of local seismic events in the area of Valaam's island}

\author[iept]{M.~A.~Nikitina\corref{corr1}}
\ead{margaritnikitina@yandex.ru}
\address[iept]{Institute of Earthquake Prediction Theory and Mathematical Geophysics, Russian Academy of Sciences, Moscow, 119991, Russia}
\cortext[corr1]{Corresponding author}

\author[spbpu,spbu]{A.~Y.~Val'kov}
\ead{alexvalk@mail.ru}
\address[spbpu]{Department of Mathematics, Peter the Great St. Petersburg Polytechnic University, St. Petersburg, 195251, Russia}
\address[spbu]{also at: Department of Physics, St.~Petersburg State University, Petrodvoretz,  St.~Petersburg, 198504, Russia}

\begin{abstract}
A lot of seismic phenomena with small magnitude has been taking place in the Ladoga lake region. One of them is the earthquake which was recorded on the 31$^\text{th}$ of July, 2010 followed by an earthquake swarm near Nikonovsky cape. There is only one station in the region that is why it is difficult to process the data.
In this work we propose a new method for obtaining main event focal mechanism and constructing the synthetic seismograms for this event.
The data of swarm events is used to test the method of constructing synthetic seismograms and the data of close events to determine the parameters of the media.
Since the amplitude of the wave P is close to zero, the location of the nodal plane can be submitted.
That is the reason why we can find the focal mechanism for the earthquake using data of the single seismostation.
Resulting fault plane has the same inclination as known tectonics line which the earthquake source is placed on.
Our proposed method is based on Green function method with using several kinds of sources and temporal $\delta$-shaped sequences.
\end{abstract}


\end{frontmatter}

\section*{Introduction}
The focal mechanism of an earthquake is one of the most important parameters characterizing a seismic event.
Nowadays it is most often calculated from the records of two or more seismic stations.
For this calculation, mainly, we need data where it is possible to clearly distinguish the first-motion direction of longitudinal P waves, or it is possible to determine the ratio of the amplitudes of different types of P and S waves.
However a dense network of seismic stations isn't everywhere. On the other hand earthquakes are not detected by several station everytime. For example, if their magnitude is very small then it can be detected only by the nearest receivers.
In our case seismic network on Ladoga lake is limited by the only one station~\citep{KA} and the magnitude of the earthquake is small, that is why the stations that are located farer, did not register any waves.

Before the main event on the 29$^\text{th}$ of July, 2010, there was a strong atmosphere pressure variation in the area of Ladoga lake and after that the seiche began.
This phenomenon is accompanied by appearance of a standing wave of a large period in the pond. In other words the enclosed body of water resonances as a whole part.
When wind ends and atmosphere pressure turns to be normal then water returns to equilibrium which occurs in the form of damped oscillations.
The height of the seiche on large lakes is usually $20$--$30$~cm.
This event (according to the authors of the article~\citep{AKN}) served as a trigger for a whole earthquake swarm of seismic events near Valaam's island. The earthquakes were detected untill the 31$^\text{th}$ of July, 2010.

\begin{figure}[ht]
\center
\includegraphics[width=0.98\textwidth]{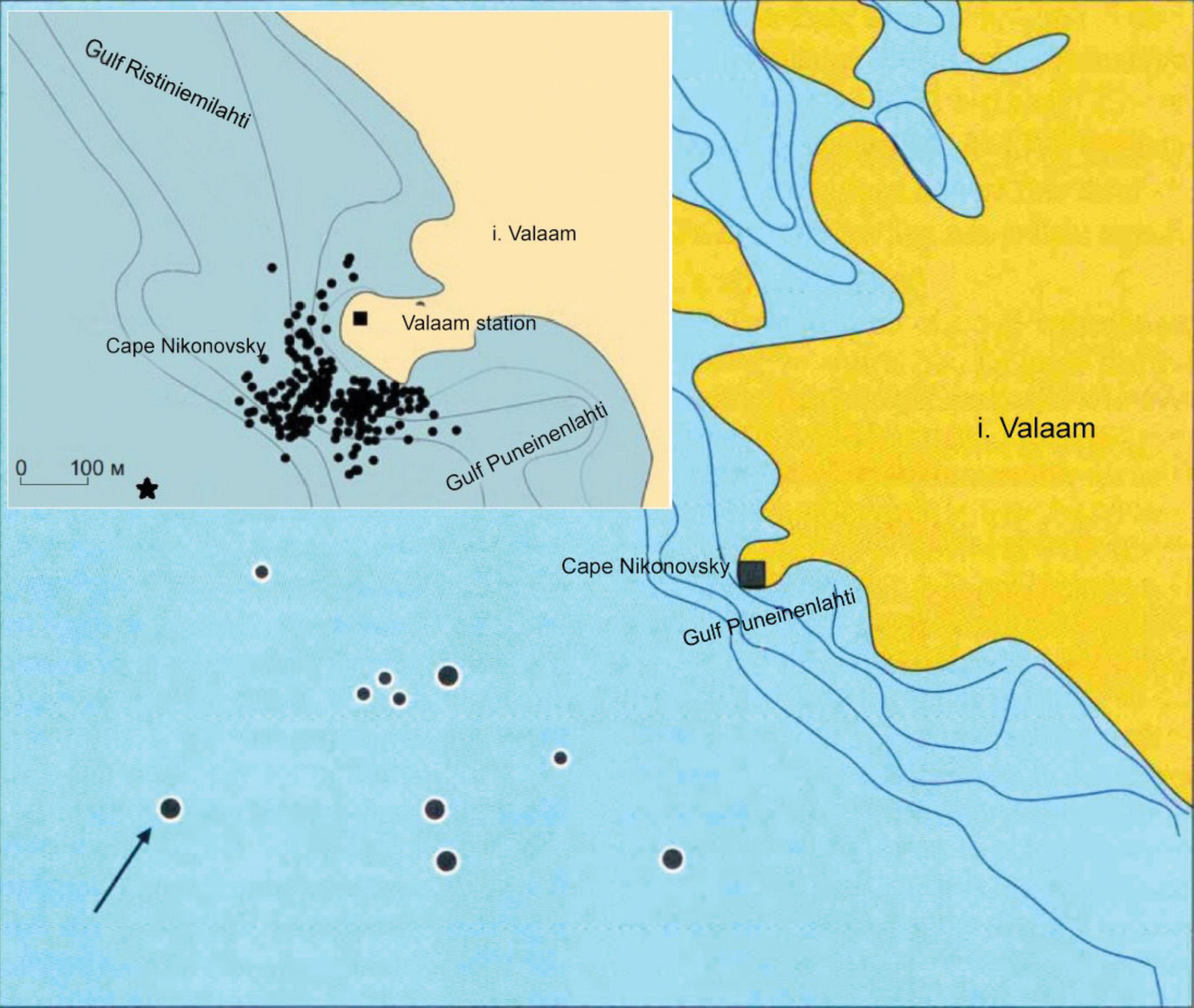}
 \caption{The map of the part of the Ladoga lake where we marked the location of the seismic station ($\blacksquare$) and several main local events with the main seismic phenomena~\citep{AKN, AOKM}. More details are in the text. } 
 \label{map}
\end{figure}

For convenience we will mark the position of the station on the map and several main local events (Fig.~\ref{map}).
We divide them into two parts, which was in 2006~\citep{AOKM} and in 2010~\citep{AKN} respectively.
The first part of events is marked with black dots on the insert in Fig.~\ref{map}. These events are taking place in the immediate close of the Nikonovsky cape.
The second part of events is the earthquake swarm ($\bullet$). In this work we will describe one of the small local events of the earthquake swarm and it is marked with a star ($\bigstar$). At the same time we are interested in the earthquake (it will be sigh by an arrow).
It is known that all the events occur in quite dense medium (crushed or monolith gabbro-diabase)~\citep{p1998}. That is why we can use nearest events to check the optimal model of medium and type of source.

The main aim of this article is to develop a method for solving such problems and as a result we will obtain the focal mechanism of the earthquake.
This work will be divided into three parts.
In the first part, we will describe the features of the proposed method for constructing a synthetic seismograms.
In second part we will demonstrated the construction of a synthetic seismogram for an event of the earthquake swarm without specifying the parameters of the source. After that we will return to the case of the main event in part three.
We will find the parameters of the earthquake source, by comparing the synthetic seismogram with the practical one. In final we will construct it's focal mechanism.
In conclusion we discuss the obtained model and the possibility of its application to other cases.

\section{Specific features of method for constructing a synthetic seismograms}

Perturbations in an elastic medium for an infinite continuum is described by the equation of motion. It can be written as~\citep{landau}:
\begin{equation}\label{1.1}
    \rho \frac{\partial^2 u_\alpha} {\partial^2 t}=\frac{\partial \sigma_{\alpha\beta}}{\partial r_\beta}+ F_\alpha,
\end{equation}
where $\u(\r,t)$ -- the displacement at point $\r$ in the time $t$, ${\bf F}$ -- the body force per unit volume, $\rho$ -- density and $\hat\sigma$ -- stress tensor.

For convenience in our case we will use the Green function method to find the displacement~\citep{k2009,v2012}.
The Green function $\hat G_0(\r,\r';t,t')$ is the field for a point source of harmonic os\-cil\-la\-tions with frequency $\omega$, which is located at the point $\r'$. For an stationary homogeneous medium it can be rewritten as $\hat G_0(\r-\r';t-t')$ and the Green function satisfy an differential tensor equation~\citep{aki}:
\begin{equation}
\label{1.12}
    \left[\rho~ \delta_{\alpha\gamma}\frac{\partial^2}{\partial t^2}- C_{\alpha\beta\gamma\zeta}\frac{\partial^2}{\partial
    r_\alpha \partial r_\zeta}\right] G_{\gamma\eta}
    (\r-\r';t-t')
    =\delta_{\alpha\eta}\delta(\r-\r')
    \delta(t-t'),
\end{equation}
where $C_{\alpha\beta\gamma\zeta}$ is a fourth-order
elasticity tensor. For an infinite isotropic media we have:
\begin{multline}\label{1.15}
    \hat G (\R,t)= \frac{1}{4\pi R^3}\left(3\frac{\R\otimes\R}{R^2}-\hat I\right)t
      {\bf 1}_{\left[R/c_\text{s},\, R/c_\text{p}\right]}(t)\\
    +\frac{1}{4 \pi R c_\text{s}^2}\left(\hat I-\frac{\R\otimes\R}{R^2}\right)\delta\left(t-\frac{R}{c_\text{s}}\right)
    +\frac{1}{4\pi R c_\text{p}^2}\frac{\R\otimes\R}{R^2}\delta\left(t-\frac{R}{c_\text{p}}\right),
\end{multline}
where $\R=\r-\r'$, $\hat I$ -- the unit tensor, and $\otimes$ -- the tensor product symbol, $c_\text{s}$ and $c_\text{p}$ -- wave S and P velocity, respectively, ${1}_{[a,b]}(t)$ is the indicator function of the interval $[a, b]$.
After that we can obtain expression for calculation of the displacement through $\hat G(\r-\r_1; t-t_1)$ and the source function:
\begin{equation}\label{4.1}
u_{\gamma}(\r,t)=u_0(\r,t)
+\int \mathrm{d}\r_1 \int^t_{-\infty}\mathrm{d}t_1 G_{\gamma \eta}(\r-\r_1;t-t_1) F_{\eta}(\r_1,t_1),
\end{equation}
where $u_0(\r,t)$ is a solution of homogeneous equation~\eqref{1.1} with ${\bf F}={\bf 0}$. In our case $u_0(\r,t)=0$ because of waves are generated only by force and at infinity waves are tend to zero.

In this work we will use three kinds of sources~\citep{Pujol209,2}
\begin{itemize}
\item[\textbf{a.}] \textbf{Dipole without a moment.}
This type of source is equivalent to fault shift.
It's formed by two equal in magnitude and oppositely directed simple point forces. They applied to the points which are located on a the line that is direction coincides with the direction of forces.

\item[\textbf{b.}]  \textbf{Double dipole without a moment.}
This type of source describes the source of an earthquake of shear type.
It's formed by two perpendicular dipoles without a moment, or by a double-couple of forces.

\item[\textbf{c.}] \textbf{Spherical source.}
This type of source can describe a spherically symmetric field of a longitudinal wave or an explosion.
Its model is the superposition of three mutually perpendicular dipoles without a moment.

\end{itemize}
As we can notice the base element for all three types is point force. Below we will detailed the expression for the displacement from this source.

Let accept the point force which changing with time $\propto X_0(t)$ and attaching to the origin along unit vector ${\bf e}_F$:
\begin{equation}\label{function}
{\bf  F}(\r_1, t_1) ={\bf e}_F X_0(t_1)\delta(\r_1).
\end{equation}
To obtain expression of displacement we substitute force function~\eqref{function} and the Green function ~\eqref{1.15} in Eq.~\eqref{4.1}. The result is the classical Stokes solution.  
In what follows we will restrict ourselves to the case of far-zone. Then the first term in Eq.~\eqref{1.15} vanishes and the displacement can be written as:
\begin{equation}\label{4.3}
u_{i}(\r,t)\simeq\frac{\delta_{ij}-\gamma_i\gamma_j}{4 \pi \rho c_\text{s}^2 r}X_0\left(t-\frac{r}{c_\text{s}}\right)
+\frac{\gamma_i\gamma_j}{4 \pi \rho c_\text{p}^2 r}X_0\left(t-\frac{r}{c_\text{p}}\right).
\end{equation}
where $\gamma_i={r_i}/{r}$ -- directional cosines and the direction of $\gamma_j$ is associated with vector ${\bf e}_F$ (cf.~\citep{aki}).



It should be noted that the real source has an impulse character extended in time, i. e. the function $X_0(t)$ is different from the $\delta$-function.
In modeling, we will use two $\delta$-shaped sequences:
the first one is based on the Lorentz function (or ``Poisson kernel''), and the second one -- on the ``Dirichlet kernel'':
\begin{eqnarray}
\label{delta1}
&&\delta_L\left(t-\frac{r}{c},\varepsilon\right)=\frac{1}{\pi}\cdot\frac{\varepsilon}{\left(t-{r}/{c}\right)^2+\varepsilon^2},\\
\label{delta2}
&&\delta_D\left(t-\frac{r}{c},\varepsilon\right)=\frac{1}{2\pi}\cdot \frac{\sin(\varepsilon^{-1}(t-{r}/{c}))}{\sin((t-{r}/{c})/2)}
\end{eqnarray}
with $\delta_{D,L}(x,\varepsilon)\to \delta(x)$ for $\varepsilon \to 0$.
The parameter $\varepsilon$ is characteristic temporal width of quasi $\delta$-functions $\delta_L(t-{r}/{c}, \varepsilon)$ and $\delta_D\left(t-{r}/{c}, \varepsilon\right)$.
It is convenient to use the Lorentz function if the resulting wave is a highly concentrated (in the time-domain) signal, and the Dirichlet kernel when we are dealing with a time-stretched and oscillated signal.

\section{Synthetic seismogram of the small local event and the source model}

We would describe the small local event which was marked by $\bigstar$ on the map in Fig.~\ref{map}.
It is known from the seismic data that the source of the event is located at a distance of about 320~m from the station and at a depth of about 100~m.
For convenience, we will use the coordinate system associated with the source. As the origin of our Cartesian system we will use the hypocentre and will direct the $x$ axis to the north (ns), and $y$ axis to the east (ew).

\begin{figure}[ht]
\center
\includegraphics[width=0.8\textwidth]{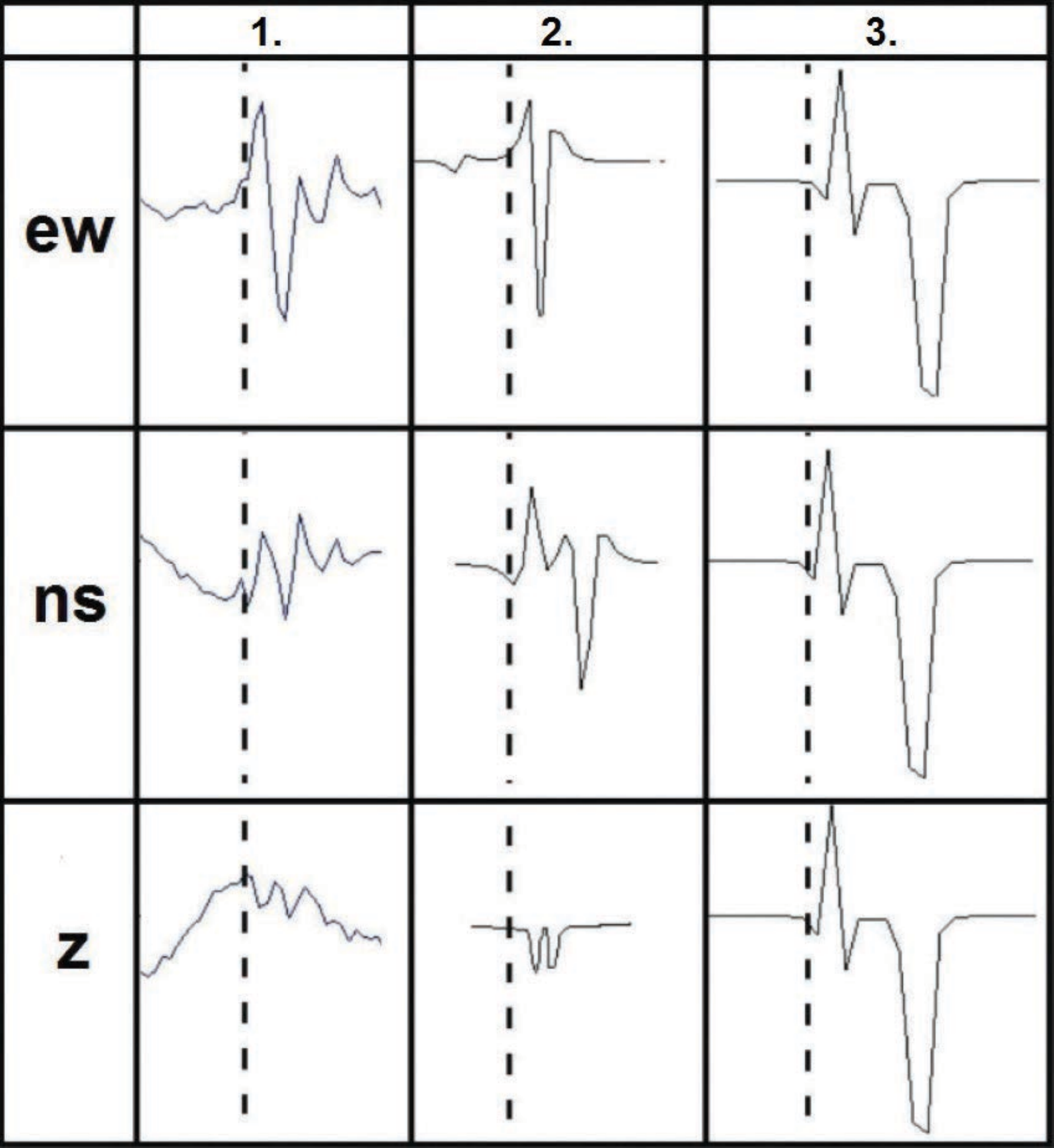}
 \caption{ The seismogram of real event (1) and synthetic seismograms for the source model of a dipole without a moment (2) and the source model of the spherical source (3). Wave S arrival time is shown by dashed line on each graphic.
}
\label{pichok}
\end{figure}

On this event we are going to test method for constructing synthetic seismograms. Known data are limited with the seismogram from a single station and there is no wave P in seismogram (the first column in Fig.~\ref{pichok}). That is why we can not localize the source or obtain a focal mechanism. Nevertheless we can make some conclusions about nature of the source. Particularly practice signal is sufficiently localized in time. Therefore we will use Lorenz function for $\delta$-function approximation~\eqref{delta1}.

We will consider two kinds of source, the spherical source and the dipole without a moment (by analogy with~\cite{4}). As we can see from Fig.~\ref{pichok} the synthetic seismogram of the spherical source does not differ for all three components while the practice seismogram is different. Therefore, the spherical source can be excluded from consideration. In the same time seismograms of the first and second columns are close enough. That is why we can conclude that a dipole without a moment well describes the source for our small local event.


\section{The earthquake source model}

The earthquake was recorded on the 31$^\text{th}$ of July, 2010 on Ladoga lake near Valaam's island.
According to the seismogram, the source is at a depth of 2~km and at a distance from the epicenter to the station is $2.3$~km approximately.
The approximate magnitude of the event is $-0.2$~\citep{AKN}.
We will use the double dipole without a moment to describe the focal mechanism for the source of the earthquake~\citep{Kasahara}, in contrast to the swarm events.
Let's describe some qualitative differences between seismogram of the earthquake and the small local event.
\begin{figure}[ht]
\center
\includegraphics[width=0.98\textwidth]{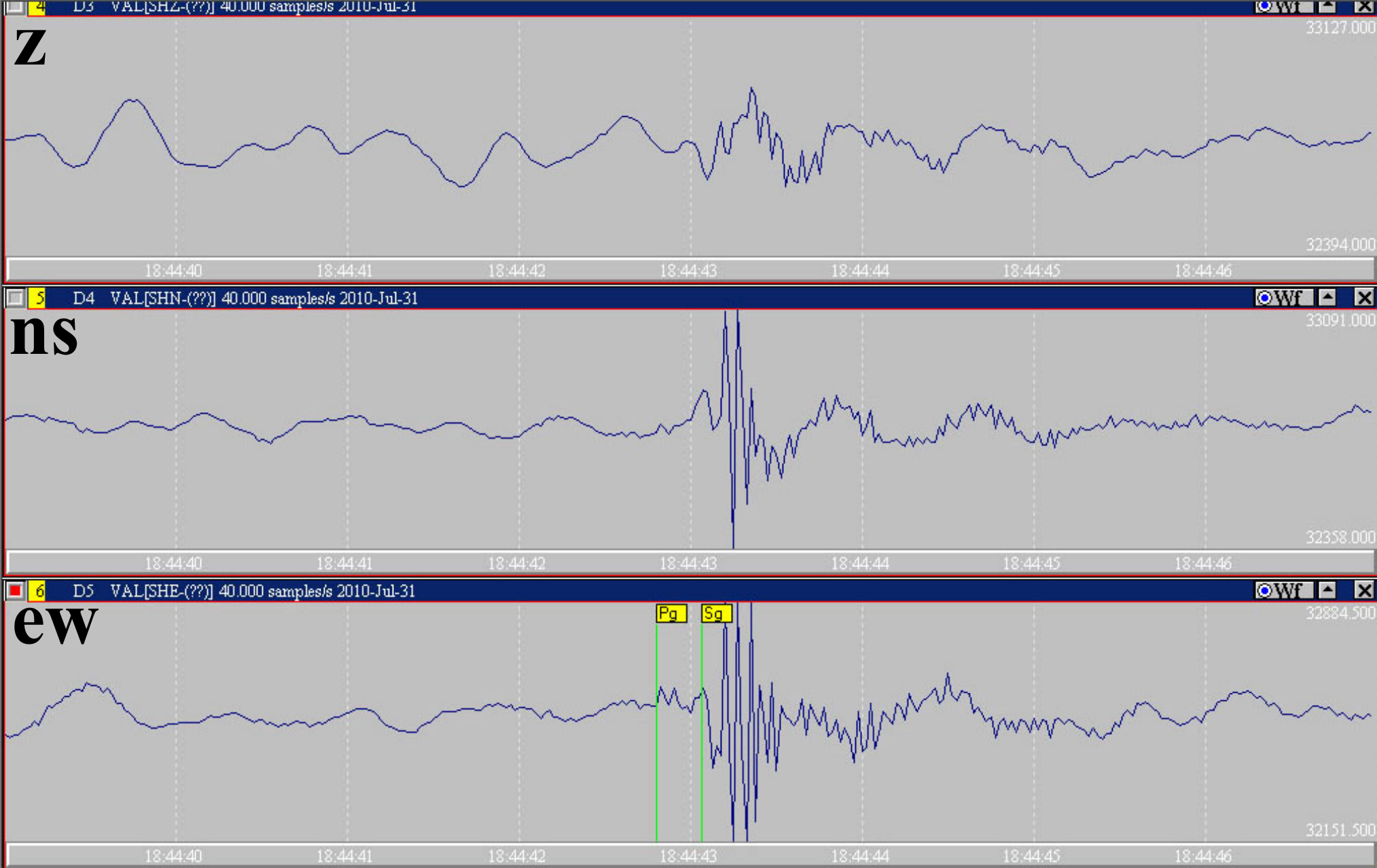}
 \caption{ The seismogram of the earthquake which was on the 31th of July, 2010 near Valaam's island~\citep{AKN}.}
 \label{pract}
\end{figure}

One of the main differences between them is the fact that the record of the earthquake spreads out in time more then for the small local even(Fig.~\ref{pract}).
That is why it is well described by the second presentation of $\delta$-function~\eqref{delta2}.
After that we will describe the features of the obtained seismogram.
We can clearly identify P wave on the ew axis, note wave P and S arrival time and get their amplitudes.
It should be noted that wave P didn't recorded on the axis ns and we can not separate P and S wave signals from the noise on the axis $z$. It means that we can suppose their absence with some level of truth.

Another difference between them is the fact that the amplitude of the wave P is very small. That is why we can locate the nodal plane. Put simplify, this feature makes possible to obtain a focal mechanism.
For the complete picture, we will write down the vectors characterizing the position of the source in space:
\begin{itemize}
  \item[]   ${\bf n}=
  \begin{pmatrix}
   -\sin\delta \sin\phi_s \\
  \sin\delta \cos\phi_s\\
  -\cos\delta \\
  \end{pmatrix}$ -- fault normal,
  \item[] ${\bf D}=
  \begin{pmatrix}
  \cos\lambda \cos\phi_s+\cos\delta\sin\lambda\sin\phi_s \\
  \cos\lambda \sin\phi_s-\cos\delta\sin\lambda\cos\phi_s\\
  -\sin\delta\sin\lambda \\
  \end{pmatrix}$ -- slip,
  \item[] ${\bf l}_{\text{SH}}=
  \begin{pmatrix}
                              -\sin\phi \\
                               \cos\phi\\
                              0 \\
  \end{pmatrix}$ -- SH-wave direction,
  \item[] ${\bf l}=
  \begin{pmatrix}
                             \sin i_{\xi} \cos\phi \\
                             \sin i_{\xi} \sin\phi\\
                             \cos i_{\xi} \\
  \end{pmatrix}$ -- P-wave direction.
\end{itemize}
We need to find several angles to construct focal mechanism: $\delta$, $\phi_s=\phi$, $\lambda$, $i_{\xi}$.
The last angle can be calculated from the location of the hypocenter and the seismostation. Namely it can be find like $\tan(i_{\xi})=x/h$, where $x$ the distance from the hypocenter to the station, $h$ -- the depth of the source.
As we said earlier the amplitude of wave P is near to zero in comparison with the S wave amplitude.
That is why we can assume location of the nodal plane and this is the reason why we can calculated other angles.

It should be also taken into account that the depth of the earthquake is significantly different from the depth of the small local event, so there is a necessity to consider damping in the medium.
We will do this by including in the formula the addition to the wave amplitude due to the quality-factor~\citep{dobr}:
\begin{equation}\label{4.8}
{\cal D}_\text{s}=\exp\left[-(\omega/2 Q_\text{s}) t\right],
\end{equation}
where $\omega$ -- the linear frequency on which the earthquake was filtered, $Q_\text{s}$ -- quality-factor for the wave S.
Thus, we add to the model frequency damping in the medium.
Summarizing all of the above,in this case we will use the formula for the double dipole without the moment~\citep{aki}. The SH-wave velocity can be written in the form:
\begin{equation}\label{4.7}
{\bf v_\text{SH}}\approx{\bf l}_\text{SH}\frac{1}{4 \pi r c_\text{s}}
\frac{d}{dt}\left[X_0(t-r/c_\text{s}){\cal D}_\text{s}\right]
\big[({\bf l}\cdot{\bf n})({\bf l}_\text{SH}\cdot{\bf D})+({\bf l}\cdot{\bf D})({\bf l}_\text{SH}\cdot{\bf n})\big].
\end{equation}

\begin{figure}[ht]
\center
\includegraphics[width=0.98\textwidth]{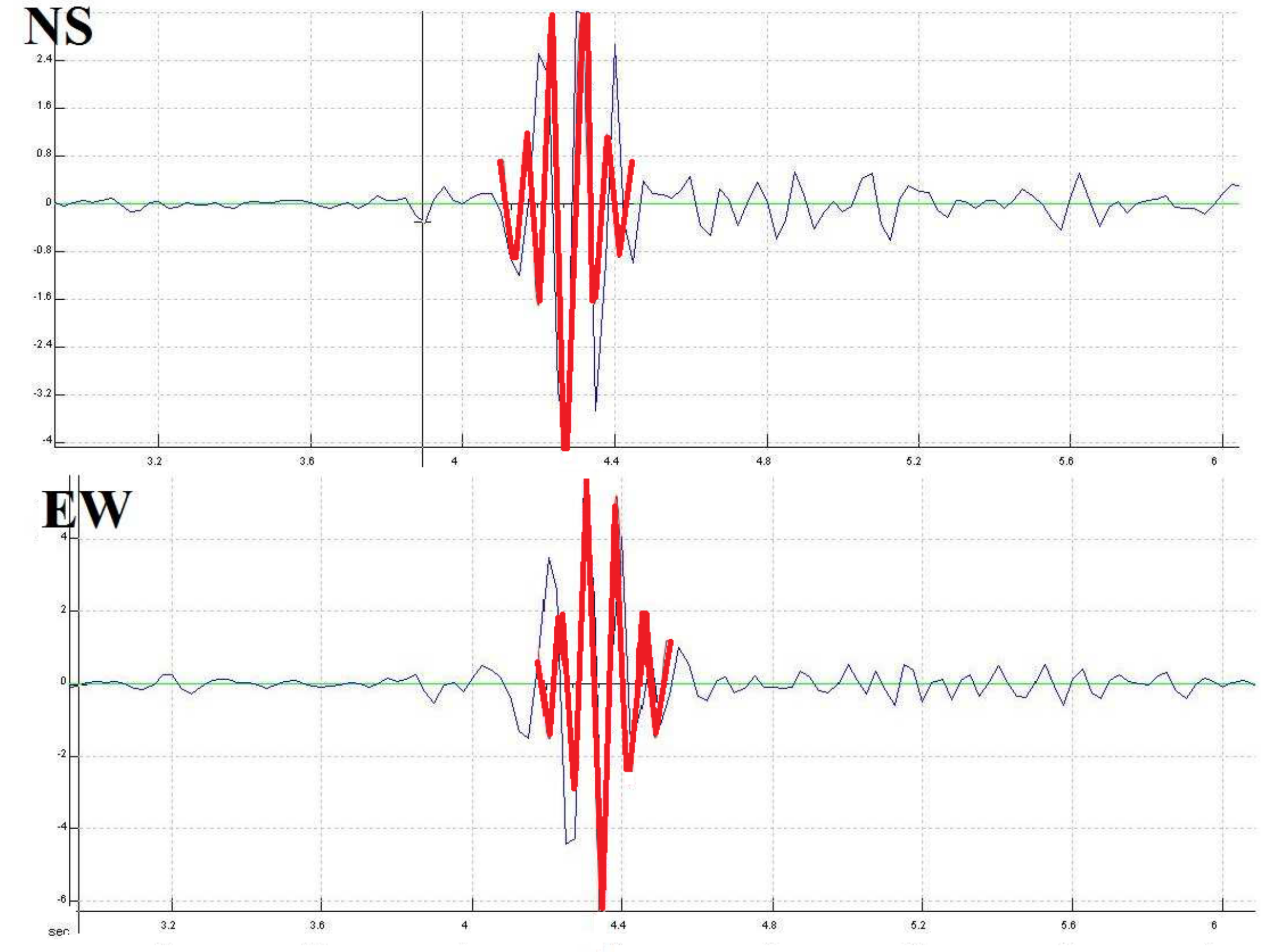}
 \caption{
 There are practice (red) and synthetic (gray) seismograms.
 They are given for two of three axes (ns and ew), since there isn't SH wave on the z axis.}
 \label{seismo}
\end{figure}

We modeling the $x$ and $y$ components of SH-wave velocity vector~\eqref{4.7}. Varying source parameters we use least squares fitting of time dependence of synthetic SH-wave velocity vector and experimental seismograms.
As the result we obtain a pretty good coincidence with a real seismogram at angles:
$$\delta=89^{\circ},\phi_s=\phi=145^{\circ},\lambda=132^{\circ},i_{\xi}=-47^{\circ},$$
which was presented below on the Fig.~\ref{seismo}.

\section{Results and discussion}

\begin{figure}[hbt!]
\center
\includegraphics[width=0.8\textwidth]{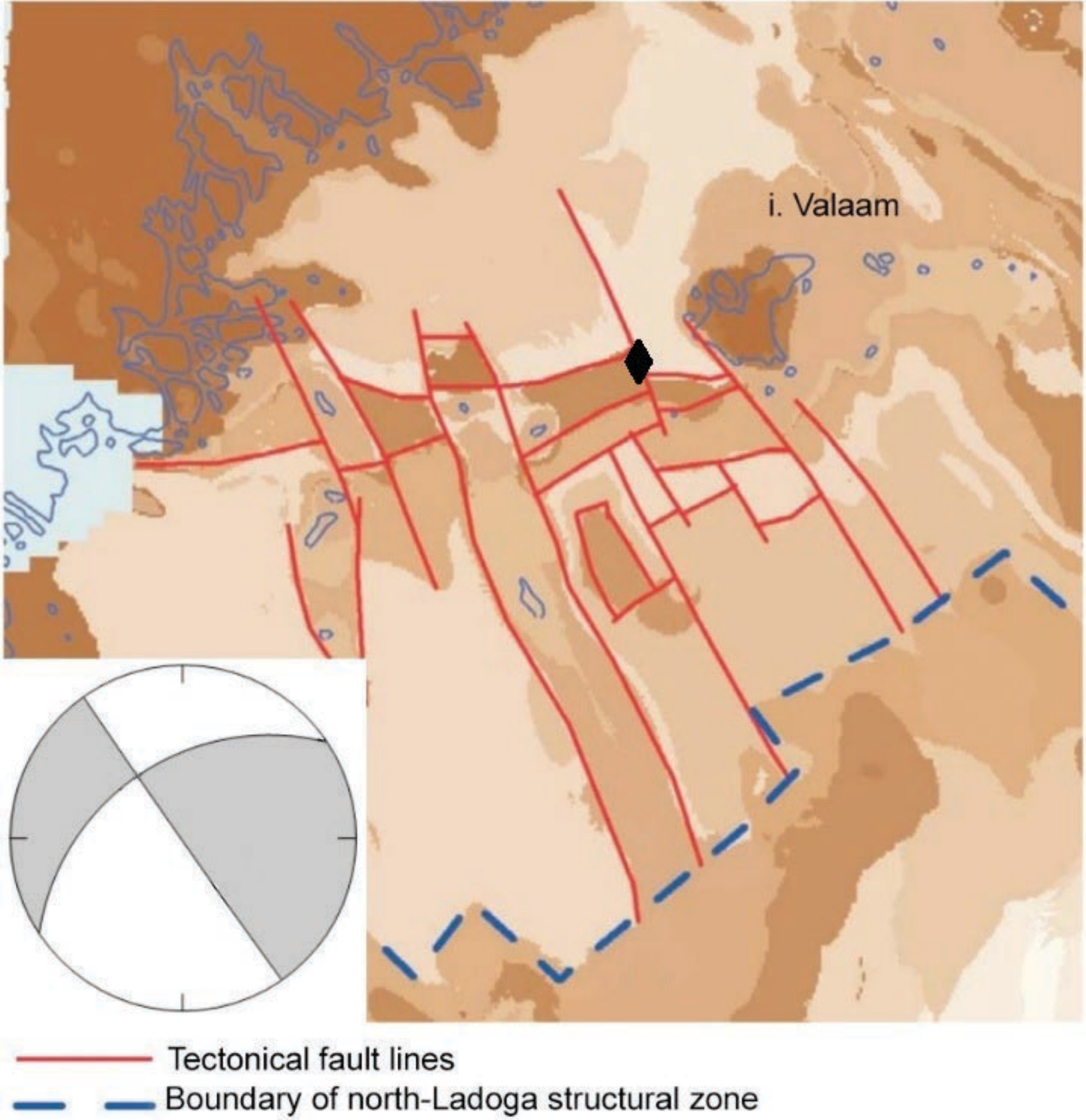}
 \caption{The map of Ladoga lake with tectonic lines on it~\citep{AOKM}. Insert shows the focal mechanism.}
 \label{tect}
\end{figure}

The major objective of this work is finding the parameters of the focal mechanism for the earthquake was achieved.
However, we can determine the accuracy of the solution only by using this method with other phenomena.
It can be noted that we can also check our solution. We can compare our focal mechanism with tectonic lines, i. e. fault plane and the tectonic line on which the earthquake source are located. Therefore tectonic lines will be shown in the map below (Fig.~\ref{tect})~\citep{AKN}. It is known from seismic data that the earthquake hypocenter locate on one of the tectonic lines $\blacklozenge$ on the (Fig.~\ref{tect}).
Here we can see that the resulting fault plane has the same inclination as known tectonics line.

For verification of our method we tested it on the problem with the known solution. We calculated the focal mechanism for the earthquake which was in 2003 in south-eastern Finland.
Seismograms for this event are known from several seismic stations and were published in~\citep{fin}.
The method proposed in our work gives the focal mechanism which differs from the result of~\citep{fin} one only by few degrees.

\section{Conclusion}

To summarize, we are interested in Ladoga region because of there are numerous insufficiently unstudied events with small magnitude. This problem is compounded by the single station in this region.
But the problem is not insurmountable. We can find the exact focal mechanism of the main event due to a close arrangement of the nodal plane by virtue of the fact that the wave P amplitude is so small. Also we have the ability to test the media model on closer events.
For our method we use several models of source and temporal delta-functional approximations for it. The displacement were calculated by Green function method with considering damping effect in media. Parameters of the source model are calculated by the best comparison of resulting synthetic and practice seismograms. After all we get that the slope of the fault plane in the focal mechanism is different only by several degrees from the known inclination of tectonic lines in this region. We assume from here that our method is very promising for solution of the similar problems.

{The work was partly supported by the Russian Foundation for Basic Research, grant No.~16-02-00465a.}
One of the authors (M.~A.~N.) is grateful to B.~G.~Bukchin for discussions and advice.

\bibliographystyle{elsarticle-harv}
\bibliography{NV_valaam}

\end{document}